**Author(s):** AKINDELE OGUNLEYE

**Title:** Exploring Study Abroad with Traditionally Underrepresented Populations: Impacts of Institutional Types

**Affiliation:** University of Texas at El Paso
Educational Leadership and Foundation
500 W University Ave, El Paso, TX 79968

**Email:** aogunleye@miners.utep.edu

**Date of submission:** June 14th, 2024.

**Exploring Study Abroad with Traditionally Underrepresented Populations: Impacts of Institutional Types**

**Abstract**

The study investigated roles of institutional types and ethnic/racial background on academic credit among the traditionally underrepresented population of the U.S. study abroad program. Using archival data, the study sampled the student's enrollment and academic credit information spanning a period of 20 years (2003 - 2022). Data analysis Using One-Way Analysis of Variance (ANOVA) indicates significant main influence of institutional type (p<.001) and significant main influence of ethnic/racial identity (p<.001) on students' academic credit. The result was discussed in terms of its relevance in educational policy re-evaluations, improvement of the study conditions of the underrepresented students, and enhancement of the enrollment opportunities of these minority population across all U.S. institutions of learning.

**Introduction**

Study abroad programs have become essential in higher education due to the world's growing interconnectedness. These programs offer students the chance to enhance their global awareness, intercultural skills, and intellectual perspectives (Engle & Engle, 2003). These programs are frequently praised for their significant influence on students, providing experiences that can profoundly influence their personal and professional lives (Lewis & Niesenbaum, 2005). The potential advantages for historically marginalized people are intriguing, as Study abroad programs can assist in closing educational and cultural disparities and provide access to novel prospects and viewpoints. Although the advantages of study abroad programs have been extensively demonstrated, there is still a lack of equal participation, especially among traditionally marginalized groups such as students of color, low-income students, and first-generation college students (Brux & Fry, 2010).

**The underrepresented population in the U.S. Study Abroad program**

Historically marginalized demographics in U.S. study abroad programs encompass various. demographic groups with lower participation rates relative to their representation in the overall student population. The following groupings are included:

1. Racial and Ethnic Minorities: Students belonging to African American, Hispanic/Latino, Asian American, Native American, and Pacific Islander communities have lower rates of participation in study abroad. According to research from the Institute of International Education (IIE), in the 2019/2020 academic year, approximately 68% of U.S. study-abroad students were white, while minority students made up a lesser percentage (IIE, 2021).

2. First-Generation College Students: Students who are the initial members of their families to enrol in college frequently encounter financial and informational obstacles that restrict their involvement in study abroad programs. They may not have the necessary familial support structures and financial resources to enable such experiences (Sweeney, 2020).

3. Students who have disabilities, including physical, learning, and psychological disorders, may face substantial difficulties when studying abroad. These challenges can range from problems with accessibility to the requirement for specialized support services (Mobility International U.S.A, 2021).

4. Low-Income Students: Financial limitations pose a significant obstacle for numerous. students contemplating participating in a study abroad program. Economically disadvantaged students may need help covering the supplementary expenses related to studying abroad, such as transport, visas, and increased living costs (NAFSA, 2020).

5. LGBTQ+ students may be dissuaded from studying abroad due to apprehensions regarding their safety, acceptance, and potential legal complications in host nations. These students may need help searching for secure and hospitable venues (IIE, 2021).

6. STEM Majors: Students in Science, Technology, Engineering, and Mathematics (STEM) fields often have rigid course requirements and limited flexibility in their academic schedules, making it difficult to fit in a study abroad experience (Institute of International Education, 2017).

The lack of representation of traditionally marginalized people in study abroad programs is a complex issue driven by multiple impediments, including financial limitations, limited access to information, and perceived cultural and familial obstacles (Kasravi, 2009; Salisbury et al., 2009). Nevertheless, it is crucial to acknowledge the impact of several institutional categories, such as community colleges, minority-serving institutions (MSIs), and research universities, on forming study-abroad involvement. Although there are now differences in study-abroad participation among traditionally underrepresented populations, there is a substantial opportunity for reform and enhancement. Studies indicate that other factors contribute to this inequality, such as financial limitations, insufficient access to information, and academic apprehensions (Salisbury et al., 2018). Moreover, the accessibility and supportiveness of study abroad programs for minority students may be influenced by institutional factors, such as size, location, and available resources (Engle & Engle, 2018). Gaining insight into the varying effects of different types of institutions on study-abroad experiences is essential for improving access and fairness in global education programs. It can stimulate beneficial transformation in higher education and promote diversity and inclusion.

**Institutional Types and Their Role**

Within U.S. study abroad programs, universities can be divided into two main categories: Associate and other higher education institutions. These groupings exhibit notable distinctions.

**Associate Institutions:** Associate institutions, also known as community colleges, primarily confer associate degrees, typically two-year degrees such as Associate of Arts (AA), Associate of Science (AS), and Associate of Applied Science (AAS). These colleges often prioritize the following areas, according to the U.S. Department of Education (2024):

1. Transfer Programs: Numerous Associate schools exhibit a significant proportion of students who successfully transfer to four-year colleges to pursue a bachelor's degree.

2. Vocational Programs: These institutions also provide vocational programs tailored to equip students with the necessary skills and knowledge for direct employment in specialized domains, such as healthcare, technical crafts, and applied sciences.

3. Community colleges offer greater affordability and accessibility in higher education due to their comparatively lower tuition prices and less stringent admission procedures when compared to four-year universities.

**All other institutions:** Other tertiary education establishments encompass four-year colleges and universities, which provide undergraduate, graduate, and doctorate programs. These institutions can be classified as public and private institutions (Academic Influence, n.d.; Top Universities, 2020):

1. Public Institutions: These universities, which state governments financially support, typically have a larger size and a more comprehensive range of resources. Some examples of universities are the University of South Carolina and University of North Carolina, Chapel Hill. They frequently provide diverse undergraduate, graduate, and professional programs.

2. Private Institutions: Private institutions, which rely on tuition, endowments, and contributions for funding, typically have higher tuition prices but provide significant financial aid opportunities. Notable examples are Harvard University and Furman University.

It is essential to acknowledge that there are substantial differences in study-abroad opportunities among these institutions (Furman University, 2024):

**Four-year Institution:** These institutions often have established study abroad programs that provide a range of options in terms of destinations and durations, including summer programs and year-long exchanges. Universities frequently maintain specialized offices and employ dedicated personnel to support students in organizing their study abroad endeavours. Additionally, they integrate study-abroad opportunities into the overall curriculum and extracurricular programs (Top Universities, 2020).

**Associate Institutions:** Community colleges' Study abroad programs are uncommon and typically less comprehensive. These programs generally are shorter and more specialized, focusing on specific subjects of study or cultural immersion experiences. Community colleges prioritize local or regional education and direct professional preparation, allocating fewer resources to international programs (Academic Influence, n.d.). Therefore, Four-year colleges offer comprehensive and diverse study abroad programs as part of a broader educational experience. In contrast, associate institutions provide more localized and career-oriented education with fewer study-abroad opportunities. When choosing between these institutions, assess how well they match a student's educational and career objectives, financial situation, and desired study-abroad experience (U.S. Department of Education, 2024).

Hence, various. types of establishments have a crucial impact on determining the rates of study-abroad participation among marginalized people. Community colleges, minority-serving institutions (MSIs), and research universities each have unique qualities that impact the accessibility and attractiveness of study-abroad programs (Raby, 2005; Laanan, 2010). Community colleges, which frequently cater to a wide range of students, including many from marginalized communities, have distinctive obstacles in encouraging participation in overseas study programs. The challenges encompass restricted financial resources, abbreviated academic programs, and an emphasis on vocational instruction (Raby & Valeau, 2007). Nevertheless, several community schools have successfully created inventive initiatives to enable study-abroad opportunities, including expeditions conducted by faculty members for shorter durations (Raby, 2005). MSIs, such as Historically Black Colleges and Universities (HBCUs), Hispanic-Serving Institutions (HSIs), and Tribal Colleges and Universities (TCUs), have a distinct advantage in assisting underrepresented students in participating in Study Abroad programs (Sweeney, 2013). These institutions frequently prioritize cultural relevance and community support, which can help reduce participation obstacles (Gasman, 2013)

Typically, large research universities possess more excellent resources and well-established Study Abroad programs. Nevertheless, the student populations in these institutions frequently lack diversity, which can lead to feelings of isolation or lack of support among marginalized students (Brux & Fry, 2010). Several measures are being taken to enhance diversity in study abroad programs at these universities, such as offering specific scholarships, conducting outreach programs, and establishing relationships with Minority-Serving institutions (MSIs) (Dolby, 2007). To comprehend the effects of different institutional types on the involvement of traditionally underrepresented communities in Study Abroad programs, it is necessary to develop a theoretical framework that combines social and psychological aspects. Several theories have been implicated, including the Theory of Planned Behaviour, Social Capital Theory, and Cultural Capital Theory. This study investigated the function of Critical Race Theory as a theoretical framework.

**Critical Race Theory (CRT)**

Critical Race Theory (CRT) provides a framework for analyzing the structural disparities and power hierarchies that impact marginalized groups' educational experiences (Delgado &

Stefancic, 2017). Critical Race Theory (CRT) highlights the significance of race and racism in influencing educational opportunities and results. It explains the reasons behind the discriminations and exclusions that reflect in the success stories of students who passed through educational institutions. There are two basic foundations upon which this explanation are erected:

1. **Institutional Racism:** CRT emphasizes the role of institutional policies and practices in maintaining racial disparities. This encompasses the presence and ease of access to study abroad programs, which may be more restricted for students at Minority-Serving Institutions (MSIs) or institutions with fewer resources. It explains the reason for the gap in students' enrollment across diverse ethnic/racial backgrounds despite the call for cultural inclusivity and justice in global educational initiatives. By examining institutional racism, Critical race theory (CRT) provides a necessary framework for understanding and confronting the complexity of race and racism at present.

2. **Counter-Storytelling:** Critical Race Theory (CRT) promotes counter-storytelling to emphasize marginalized communities' experiences and viewpoints. By magnifying marginalized students' perspectives, institutions can better comprehend and effectively tackle the obstacles they encounter when pursuing international education. While this serves as a way to study the plights of students of minority identity, it provides important tools for advocating for social justice and fairness, including the ability to challenge prevailing narratives and emphasise the structural components of racial inequity. It also provides a means of understanding the necessary steps to take in tackling the menace of educational discrimination and injustice.

## Participation Rates and Demographics

Empirical research emphasizes the ongoing lack of minority students in study abroad programs. According to the 2021 Open Doors Report of the Institute of International Education (IIE), students of color constitute over 40% of the college population in the United States. However, only around 30% of study-abroad participants belong to these groups. Research reveals multiple obstacles that lead to this inequality. Financial limitations provide a substantial impediment, as numerous. marginalized students require assistance covering the supplementary expenses of studying abroad (Salisbury et al., 2009).

Furthermore, there is a requirement for increased knowledge and dissemination of information regarding the existing opportunities to facilitate involvement (Kasravi, 2009). Research indicates that the level of assistance institutions provide substantially impacts the rates at which minority students choose to study abroad. For instance, research has demonstrated that scholarships and fellowships explicitly targeted towards minority students positively increase participation rates (Stroud, 2010). In addition, educational establishments that offer pre-departure orientations, peer mentoring, and culturally appropriate programming can mitigate worries and improve the study abroad experience for these students (Brux & Fry, 2010).

Case studies from multiple institutions exemplify effective techniques for enhancing study-abroad participation among disadvantaged communities. For example, Historically Black Colleges and Universities (HBCUs) have implemented initiatives integrating service learning and community involvement, aligning with their students' beliefs and preferences (Gasman, 2013). In a similar vein, certain community schools have formed collaborations with foreign universities to establish Study Abroad opportunities that are both cost-effective and easily accessible (Raby, 2005). Studies have shown that minority students participating in Study Abroad programs have notable advantages, such as higher academic achievement, heightened cultural understanding, and better job opportunities (Sweeney, 2013;

; Brux & Fry, 2010).    These results highlight the significance of increasing the availability of study-abroad opportunities to all students.

**The present study**

Research on study abroad indicates that it can improve post-secondary achievement, retention, and completion (Engel, 2017). Previous research on the impacts of study abroad has often concentrated on specific institutions, such as the University of Minnesota-Twin Cities, the University of California-San Diego, and the University of Texas-Austin. However, there have also been studies that have looked at entire state systems, such as Georgia and Florida (Engel, 2017). In summary, this research has shown that those who participated in study-abroad programs had more excellent retention and graduation rates than those who did not study abroad. An investigation conducted by Posey (2003) examined the duration of degree programs and the rate of degree completion for participants in study-abroad programs within the Florida State System. The study focused on Associate, bachelor, and graduate degree programs. Posey's findings revealed that 93% of students who participated in study abroad programs successfully obtained their degrees, in contrast to only 64% of students who did not participate in such programs. The impact was particularly evident in four-year bachelor's programs, with 81% of students who studied abroad completing their degrees, compared to only 57% of those who did not participate in such programs. Based on a study conducted by Barclay-Hamir (2011) using student data from 2002 at the University of Texas-Austin, it was discovered that students who participated in studying abroad had a 60% probability of graduating within four years, whereas non-participants had a 45% probability.

Based on data from the University of California at San Diego (2010), persons who did not study abroad had a graduation percentage of 82.3% over five years. In contrast, study-abroad participants had a graduation rate of 94.7%. According to the Georgia Learning Outcomes of Students Studying Abroad Research Initiative (GLOSSARI) conducted by the University of Georgia, the most extensive study on study abroad outcomes to date, a student who is not part of a study abroad program and is enrolled full-time has a probability of less than 50% of graduating within six years. The graduation rate for students in study abroad programs is 88.7%. Luo and Jamieson-Drake (2015) found a correlation between participation in study-abroad programs and enhanced academic performance, satisfaction, communication abilities, and graduation rates. Additional research has demonstrated that participating in a study abroad program has a noteworthy influence on an individual's personal development (Dolby, 2007), ability to be self-reliant (Hadis, 2005), and belief in their capabilities (Cubillos & Ilvento, 2013). Despite the potential positive outcomes of studying abroad, there is considerable diversity in the individuals who take part, leading to the exclusion of specific segments of the student population who are expected to benefit significantly from the study abroad program (Engel, 2017).

The consistent disparity between the number of students who express their intention to participate and those who do so concerns international practitioners who aim to enhance programming and student engagement. Earlier studies have forecasted that the global health and financial circumstances caused by COVID-19 will exacerbate the existing gap even more (IIE 2020). Even once the crisis subsides, persistent health concerns and a worldwide economic downturn may necessitate fresh institutional initiatives to encourage students to participate in international experiences. Therefore, current events emphasize the necessity for investigations that enhance our comprehension of institutional categories as tools in students' decision-making and provide valuable insights that can guide practical initiatives to improve study-abroad participation (e.g., Booker, 2001; Kim & Lawrence, 2018; Peterson, 2003). In this study, I analyzed longitudinal student data from several institutions. I conducted a study to examine how different types of institutions and the racial/ethnic features of students affect their academic credit while studying abroad (Institute of International Education in 2023).

**Research Objectives:**

1. To examine the extent of study abroad participation among traditionally underrepresented populations across different institutional types.

2. To assess the academic outcomes of study abroad experiences for underrepresented students, considering institutional differences.

3. To develop recommendations for institutions to enhance support and accessibility of study abroad programs for traditionally underrepresented populations.

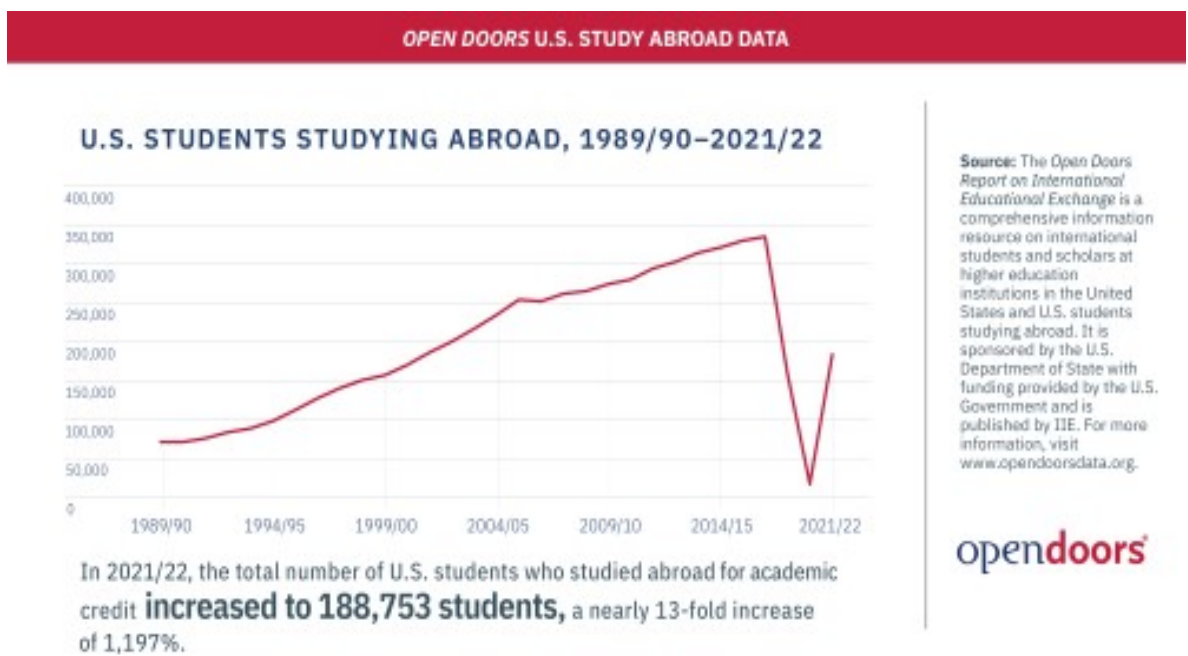

**Methods**

**Research Design**

This study employed a quantitative research design to explore the impacts of institutional types on study-abroad participation among traditionally underrepresented populations. This involved analyzing existing data sets to identify participation trends and demographic characteristics to gain deeper insights into the barriers and facilitators of study abroad participation.

**Participants**

This study includes all U.S. study abroad students who participated between 2003 and 2022. The participants were categorized based on ethnic/racial background. Thus, the study focU.S.es on examining the academic well-being of ethnic/racial minorities in the two major institutional types.

**Data Collection**

The data were obtained from the Institute of International Education's Open Doors Report (2023), which provides comprehensive statistics on study-abroad participation by demographic characteristics and institutional types. Additionally, institutional data from selected community colleges, minority-serving institutions (MSIs), and research universities were analyzed to identify patterns and disparities in study-abroad participation.

**Sample Selection**: The sample includes 6 institutions representing a diverse array of community colleges, MSIs, and research universities' activities over a 20-year period. These institutions were selected based on their geographic diversity, student demographics, and study abroad program offerings.

**Variables**: Student demographics (race/ethnicity), institutional characteristics (type), and study abroad participation rates.

**Data Analysis**

Data analysis in this study involved descriptive and inferential statistics: Means and standard deviations which were calculated to summarize the data. One-way Analysis of Variance (ANOVA) tests, using the SPSS version 26, explored the mean differences between factors in categorical variables (Institutional type: Associate versus. others, and student characteristics: ethnic/racial divides). The study ensured internal validity and reliability using well-established data sources and robU.S.t statistical methods.

**Ethical Consideration**

The Institutional Review Board (IRB) provided ethical approval before data collection.

**Results**

**Table 1**
Means and Standard Deviations of Study Abroad Student's Enrollment in 20 Years (2003-2022).

| | No of years | Mean | Std. Deviation | Std. Error | 95% Confidence Interval for Mean | | Minimum | Maximum |
|---|---|---|---|---|---|---|---|---|
| | | | | | Lower Bound | Upper Bound | | |
| Doctorate Universities | 20 | 66543.10 | 23961.28 | 5357.10 | 55328.87 | 77757.32 | 8031.00 | 97266.00 |
| Master's Colleges & Universities | 20 | 18827.15 | 7269.37 | 1625.50 | 15424.98 | 22229.31 | 967.00 | 26860.00 |
| Baccalaureate Universities | 20 | 13353.30 | 3904.38 | 873.05 | 11525.99 | 15180.60 | 1110.00 | 15904.00 |
| Associate Colleges | 20 | 842.65 | 1139.10 | 254.89 | 309.1614 | 1376.138 | .00 | 2785.00 |
| Special Focus. Institutions | 20 | 601.25 | 890.10 | 199.21 | 184.2967 | 1018.203 | .00 | 2282.00 |
| Total | 100 | 20033.49 | 26844.98 | 2684.50 | 14706.86 | 25360.11 | .00 | 97266.00 |

The results of means and standard deviations in Table 1 above show that Doctorate Universities scored higher than any other institution on study abroad student enrollment (Mean= 66543.10; SD = 23961.28), followed by Master's Colleges and universities which scored a total mean of 18827.15 (SD = 7269.37). Baccalaureate Universities Scored a total mean of 13353.30 (SD = 3904.38), while Associate Colleges scored 842.65 (SD = 1139.10). The last study abroad student enrollment is found among Special Focused Institutions, with a total mean score of 601.25 (SD = 890.10).

**Table 2**

**ANOVA Summary of the influence of institutional types on study abroad enrollments.**

|  | Sum of Squares | df | Mean Square | F | Sig. |
|---|---|---|---|---|---|
| Between Groups | 59102486166.1 | 4 | 14775621541.5 | 114.66 | .000* |
| Within Groups | 12242157260.9 | 95 | 128864813.27 |  |  |
| Total | 71344643426.9 | 99 |  |  |  |

*\*p<.001*

The ANOVA summary in Table 2 above indicates a significant primary influence of institutional type on study abroad students' students' enrollment, F (4,95) = 114.66, p < .001. This means there is a substantial difference in students' enrollments among these institutions, as shown by their means of difference. This is also a reflection of students' preferences and intentions.

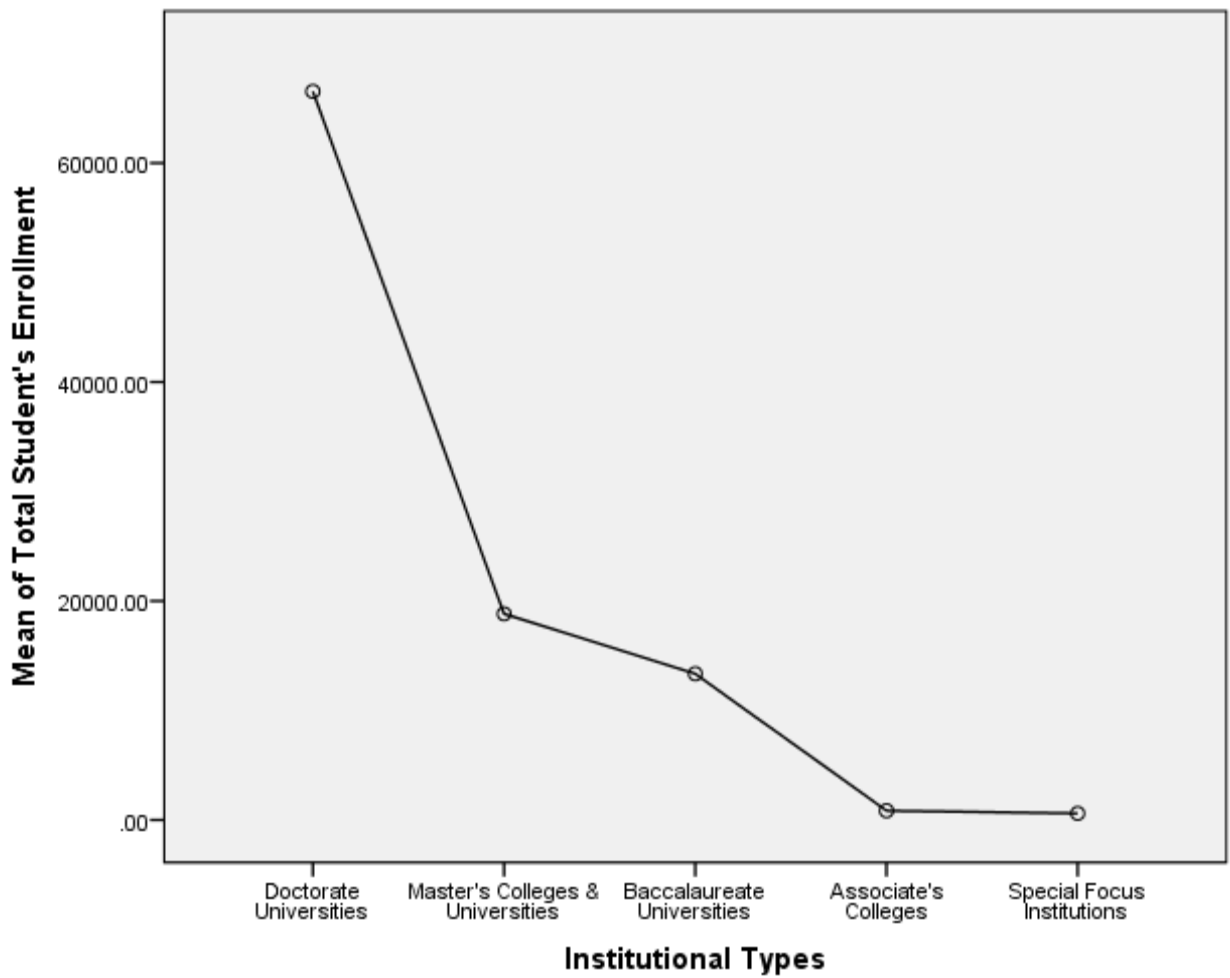

**Table 3**

**Means and Standard Deviations of Institutional Type and Student Characteristics on Academic Credit (2003-2022, no credit in 2020)**

| | | N | Mean | Std. Devia. | Std. Error | 95% Confidence | | Mini | Maxi |
| | | | | | | Lower Bound | Upper Bound | | |
|---|---|---|---|---|---|---|---|---|---|
| Asso Institution | American Indian or Alaska Native | 19 | .53 | .30 | .06976 | .3903 | .6834 | .00 | 1.10 |
| | Asian, Native Hawaiian or Pacific Islander | 19 | 4.17 | 1.45 | .33218 | 3.4758 | 4.8716 | .00 | 7.70 |
| | Black or African-American | 19 | 6.12 | 2.20 | .50490 | 5.0550 | 7.1766 | .00 | 10.10 |
| | Hispanic or Latino | 19 | 16.37 | 6.70 | 1.54 | 13.1385 | 19.5984 | .00 | 26.80 |
| | Multiracial | 19 | 3.35 | 2.26 | .51873 | 2.2575 | 4.4372 | .00 | 9.10 |
| | White | 19 | 64.18 | 17.47 | 4.01 | 55.7636 | 72.6048 | .00 | 81.40 |
| | Total | 114 | 15.79 | 23.56 | 2.207 | 11.4158 | 20.1596 | .00 | 81.40 |
| All Institution | American Indian or Alaska Native | 19 | .47 | .06 | .01289 | .4466 | .5008 | .40 | .60 |
| | Asian, Native Hawaiian or Pacific Islander | 19 | 7.72 | 1.04 | .24052 | 7.2105 | 8.2211 | 6.10 | 10.00 |
| | Black or African American | 19 | 4.88 | 1.00 | .22983 | 4.4014 | 5.3671 | 3.40 | 6.40 |
| | Hispanic or Latino | 19 | 8.16 | 2.41 | .55241 | 7.0026 | 9.3237 | 5.00 | 12.30 |
| | Multiracial | 19 | 3.01 | 1.50 | .34527 | 2.2799 | 3.7306 | 1.20 | 5.30 |
| | White | 19 | 75.76 | 5.62 | 1.289 | 73.0502 | 78.4656 | 68.3 | 83.70 |
| | Total | 114 | 16.67 | 26.80 | 2.51 | 11.6939 | 21.6394 | .40 | 83.70 |

The results of means and standard deviations in Table 3 above show that Whites obtained the higher total mean in academic credit in both associate institutions (Mean = 64.18; SD = 17.47) and All other institutions (Mean = 75.76; SD = 5.62), followed by Hispanic or Latino who scored a total mean of 16.37 (SD = 6.70) in Associate institution, and a total mean of 8.16 (SD = 2.41) in All other institutions. The lowest academic credit is found among American Indians or Alaska Natives, who scored a total mean score of 0.53 (SD = 0.30) in Associate institutions and a total mean of 0.47 (SD = 0.06) in All other institutions.

**Table 4**
**ANOVA Summary of the role of Institutional type and Student Characteristics on academic credit**

|  |  | Sum of Squares | df | Mean Square | F | Sig. |
|---|---|---|---|---|---|---|
| Asso Institution | Between Groups | 56208.483 | 5 | 11241.70 | 186.18 | .000 |
|  | Within Groups | 6521.000 | 108 | 60.38 |  |  |
|  | Total | 62729.483 | 113 |  |  |  |
| All Institution | Between Groups | 80405.606 | 5 | 16081.12 | 2312.21 | .000 |
|  | Within Groups | 751.127 | 108 | 6.96 |  |  |
|  | Total | 81156.733 | 113 |  |  |  |

*p<.001

ANOVA summary in Table 4 above indicates a significant primary influence of student characteristics (by institutional type) on study abroad students' academic credit. In Associate Institutions, ethnic/racial characteristics significantly influence academic credit, $F(5,108)$ = 186.18, $p < .001$. Similarly, in All other Institutions, ethnic/racial characteristics significantly influence academic credit, $F (5,108)$ = 2312.21, $p <. 001$. This means there is a significant difference in students' academic credit in these institutions based on their ethnic/racial characteristics, as shown by their means differences.

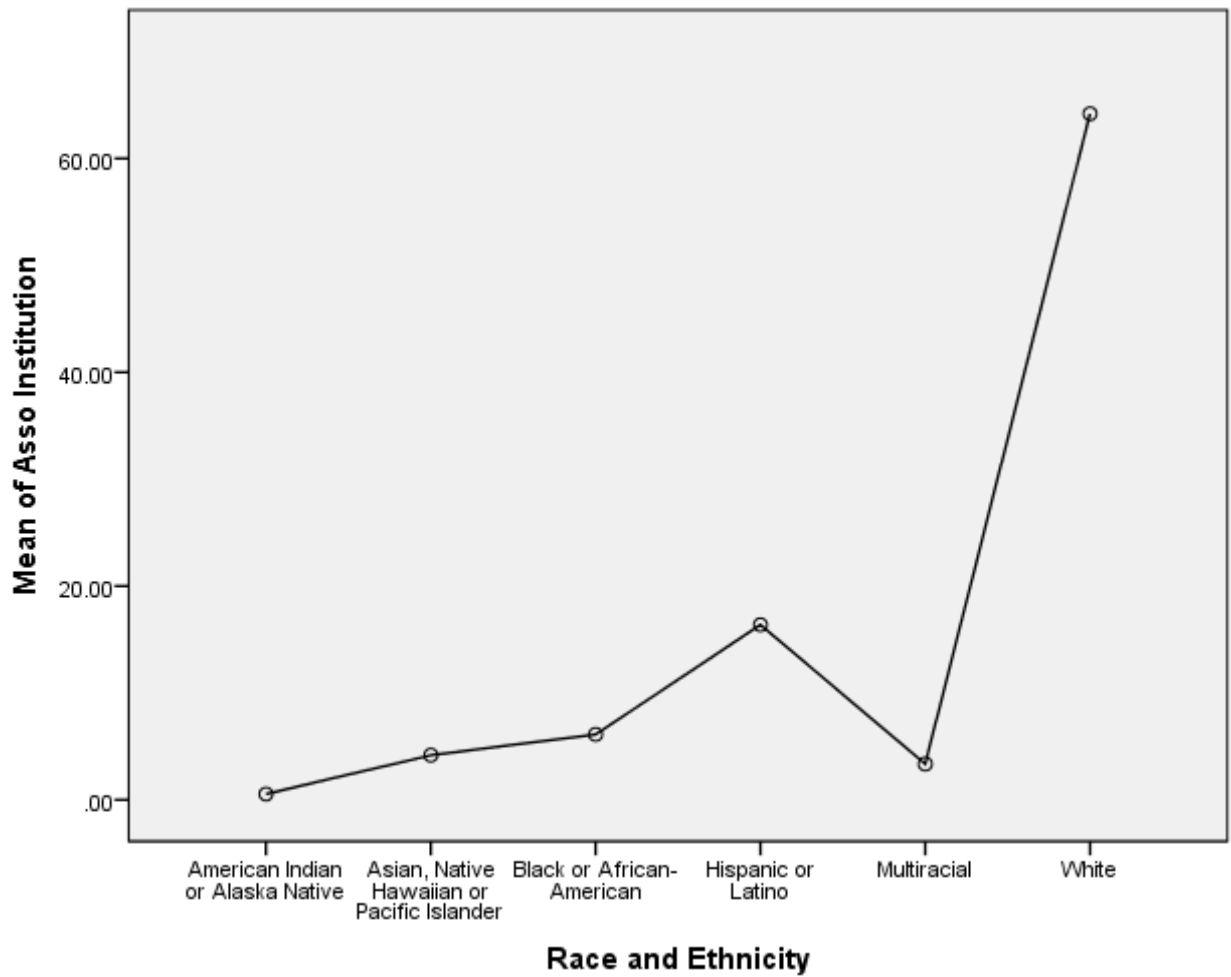

*Figure 2:* Graph showing the role of Associate Institutions and Students Characteristics on academic credit.

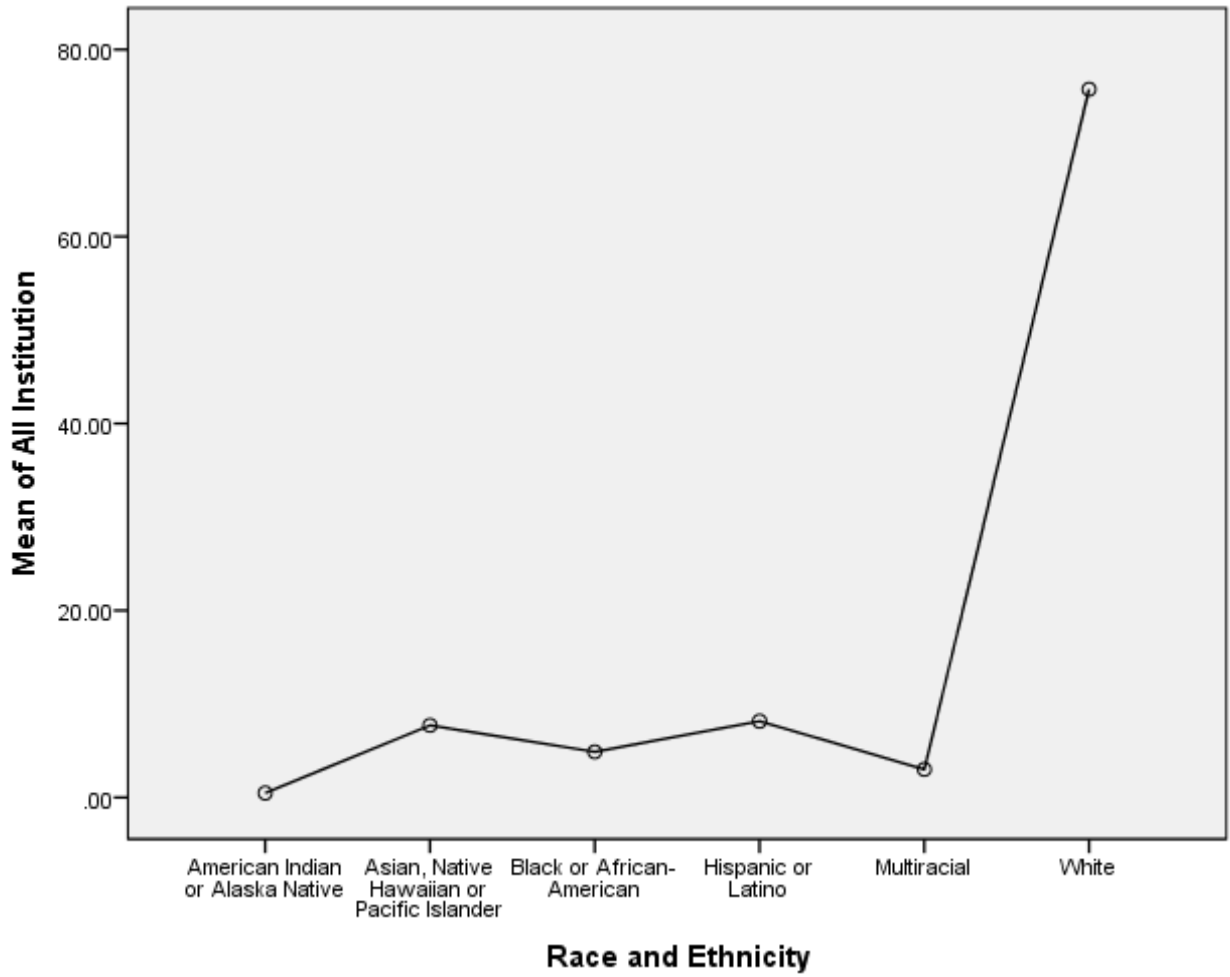

*Figure 3:* Graph showing the role of All Institutions and student characteristics on academic credit.

**Discussion**

The findings indicate that people's ethnic and racial backgrounds significantly impacted their academic credit at all institutions, even those affiliated with the Associate program. Based on the findings of the mean differences, it can be deduced that the White population was more favoured, while the American Indians or Alaska Natives were the most disadvantaged. Brux and Fry (2010), Kasravi (2009), Salisbury et al. (2009; 2018), and Engle and Engle (2018) all concluded that the underrepresented groups need more opportunities for academic progress. The findings of this study are consistent with those of the researchers above. In addition, this is not jU.S.t a reflection of the lack of opportunities available to students but also of the attention shown by the school. This may be the consequence of Critical Race Theory (Delgado & Stefancic, 2017), which emphasizes that institutional policies and practices can perpetuate racial inequality. This covers the availability of study abroad programs and their accessibility, which may be more restricted for students attending middle school institutions or schools with fewer resources. Therefore, to address the needs of these underrepresented groups, targeted initiatives that could change the institutional policies on ethnic and racial identity are required. These initiatives include scholarships, individualized support services, and inclusive program designs that reduce barriers and create more equitable access to opportunities for international education. With this information, policy and practice will be better informed, which will assist in improving fairness and access in global education projects.

**Conclusion**

The results of this study shed light on the intricate relationship that exists between the qualities of the institution and the ethnic and racial background of the students for historically underrepresented communities in terms of their academic accomplishments. This study examines the academic credits that students have earned to gain an understanding of the obstacles and opportunities that exist within various institutional settings. It has been determined from the findings of the study that racial and ethnic minorities are the most disadvantaged in terms of academic success. All schools that provide study abroad programs should see this as a wake-up call since they have developed to fulfil the intercultural inclusiveness and competency requirements of the global education programs.

**Acknowledgment**

# References

Academic Influence. (n.d.). Carnegie Classifications, College Tiers, and What They Mean. Retrieved May 25, 2024, from *https://academicinfluence.com/articles/college-rankings/carnegie-classifications-college-tiers-and-what-they-mean*

Ajzen, I. (1991). The theory of planned behavior. *Organizational Behavior and Human Decision Processes, 50*(2), 179–211.

Bourdieu, P. (1986). The forms of capital. *In J. Richardson (Ed.), Handbook of Theory and Research for the Sociology of Education* (pp. 241–258). Greenwood.

Brux, J. M., & Fry, B. (2010). Multicultural students in study abroad: Their interests, issues, and constraints. *Journal of Studies in International Education, 14*(5), 508-527.

Coleman, J. S. (1988). Social capital in the creation of human capital. *American Journal of Sociology, 94*(Supplement), S95-S120.

Delgado, R., & Stefancic, J. (2017). *Critical Race Theory*: An Introduction. NYU Press.

Dolby, N. (2007). Reflections on nation: American undergraduates and study abroad. *Journal of Studies in International Education, 11*(2), 141–155.

Engle, L., & Engle, J. (2003). Study abroad levels: Toward a classification of program types. Frontiers: *The Interdisciplinary Journal of Study Abroad, 9*(1), 1-20.

Engle, L., & Engle, J. (2018). Student perceptions of university support for study abroad at three types of institutions: Public, private and for-profit. Frontiers: *The Interdisciplinary Journal of Study Abroad, 30* (1), 30-52.

Furman University. (2024, February 21). What's the Difference Between College and Universities? Retrieved May 25, 2024, from *https://www.furman.edu/blog/whats-the-difference-between-college-and-universities/*

Gasman, M. (2013). The changing face of historically black colleges and universities. Philadelphia: *University of Pennsylvania Press.*

Gilchrist, C., & Mandlehr, K. (2009). *11reSource Spotlight.*

Gutierrez, R., Auerbach, J., & Bhandari, R. (2009). Expanding U.S. study abroad capacity: *Findings from an IIE-forum survey. Expanding study abroad capacity at U.S. colleges and universities, 6-20.*

Institute of International Education (IIE). (2021). *Open Doors Report on International Educational Exchange. Retrieved from https://opendoorsdata.org/.*

Institute of International Education. (2023). Characteristics Of Study Abroad Students For Academic Credit From Associate Institutions 2003/04 - 2021/22. *Open Doors Report on International Educational Exchange*. Retrieved from https://opendoorsdata.org/

Kasravi, J. (2009). Factors influencing the decision to study abroad for students of color: Moving beyond the barriers. *In R. Lewin (Ed.), The handbook of practice and research in study abroad (pp. 99-117). New York: Routledge.*

Laanan, F. S. (2010). Community college study abroad programs: An overview. *In M. Tillman (Ed.), A handbook for advancing comprehensive internationalization: What institutions can do and what students should learn (pp. 103-110)*. Washington, DC: American Council on Education.

Lewis, T. L., & Niesenbaum, R. A. (2005). The benefits of short-term study abroad. *Chronicle of Higher Education, 51*(39), B20.

McPherson, M. P., & Heisel, M. (2010). Creating successful study abroad experiences. *In Higher education in a global society. Edward Elgar Publishing.*

Mobility International U.S.A. (2021). "*People with Disabilities Studying Abroad." Retrieved from [MIU.S.A.org]* (https://www.miU.S.a.org).

NAFSA: Association of International Educators. (2020). "Financial Aid for Study Abroad." Retrieved from [NAFSA.org] (https://www.nafsa.org).

Raby, R. L. (2005). Internationalizing the curriculum: On- and off-campus. strategies. *New Directions for Community Colleges, 2005(131),* 57-66.

Raby, R. L., & Rhodes, G. M. (2018). Promoting Education Abroad among Community College Students: Overcoming Obstacles and Developing Inclusive Practices. *In Promoting Inclusion in Education Abroad (1st ed., pp. 114–132). Routledge. https://doi.org/10.4324/9781003446545-9*

Raby, R. L., & Valeau, E. J. (2007). Community college international education: Looking back to forecast the future. *New Directions for Community Colleges, 2007(138)*, 5-14.

Salisbury, M. H., Paulsen, M. B., & Pascarella, E. T. (2018). Why do high-impact practices work? Understanding the mechanisms of student success. *ASHE Higher Education Report, 44*(6), 1–28.

Salisbury, M. H., Umbach, P. D., Paulsen, M. B., & Pascarella, E. T. (2009). Going global: Understanding the choice process of the intent to study abroad. *Research in Higher Education, 50*(2), 119–143.

Stroud, A. H. (2010). Who plans (not) to study abroad? An examination of U.S. student intent. *Journal of Studies in International Education, 14*(5), 491–507.

Sweeney, K. (2013). Internationalizing HBCUs: Broadening the horizons for students, faculty, and institutions. *New York: Institute of International Education.*

Sweeney, K. (2020). "Barriers to Study Abroad for First-Generation College Students." *Journal of International Education Research.*

Top Universities. (2020, January). What is an Associate degree? *Retrieved May 25, 2024, from https://www.topuniversities.com/what-associates-degree*

U.S. Department of Education. (2024). Carnegie Classification of Institutions of Higher Education. *Retrieved May 25, 2024, from https://carnegieclassifications.acenet.edu/*

Whatley, M., & Raby, R. L. (2020). Understanding Inclusion and Equity in Community College Education Abroad. *Frontiers (BostonMass.), 32*(1), 80–103. https://doi.org/10.36366/frontiers.v32i1.435